\documentclass[prl,preprint,superscriptaddress,showpacs]{revtex4}

\usepackage{graphicx}
\usepackage{natbib}

\begin{document}

\title{Entanglement of a Mesoscopic Field with an Atom induced by Photon Graininess in a Cavity}
\author{A. Auffeves}
\author{P. Maioli}
\author{T. Meunier}
\author{S. Gleyzes}
\author{G. Nogues}
\author{M. Brune}
\author{J.M. Raimond}
\affiliation{Laboratoire Kastler Brossel, D\'epartement de Physique 
de l'Ecole Normale Sup\'erieure, 24 rue Lhomond, F-75231 Paris 
Cedex 05, France}
\author{S. Haroche}
\affiliation{Laboratoire Kastler Brossel, D\'epartement de Physique 
de l'Ecole Normale Sup\'erieure, 24 rue Lhomond, F-75231 Paris 
Cedex 05, France}
\affiliation{Coll\`ege de France, 11 place Marcelin Berthelot, 
F-75231 Paris Cedex 05, France}
\date{\today}

\begin{abstract}
We observe that a mesoscopic field made of several tens of
microwave photons exhibits quantum features when interacting with a single Rydberg atom in a 
high-$Q$ cavity. The
field is split into two components whose phases differ by an angle inversely
proportional to the square root of the average photon number. The field and the
atomic dipole are phase-entangled. These
manifestations of photon graininess vanish at the classical limit. This experiment opens the way to
studies of large Schr\"odinger cat states at the quantum-classical boundary.
\end{abstract}

\pacs{03.65.-w, 03.67.Mn, 42.50.Dv}
\maketitle

Known since the early days of molecular beams, the Rabi oscillation 
of a two-level atom in a coherent field plays a fundamental
role in quantum optics.
When the field is 
classical (i.e. made of a huge number of photons whose 
graininess is negligible) it is not affected
by the coupling to the atom, which oscillates between the two levels 
at a frequency proportional to the field amplitude. When
the field contains no photons, a situation relevant 
to the Cavity Quantum Electrodynamics (CQED) context 
\cite{QC_MABUCHIREVIEW02}, the Rabi oscillation 
occurs at the much slower vacuum Rabi frequency $\Omega$,
proportional to the zero-point field fluctuations in the cavity. The 
field, oscillating between the $0$ and
$1$ photon states, is then strongly affected by the coupling. The 
Rabi oscillation results in periodic maximum atom-field entanglement \cite{ENS_RMP}
and in other various atom-field effects either in the optical
\cite{QC_OROZCOCORREL98,QC_KIMBLEMICROSCOPE00,QC_REMPEFEEDBACK02} 
or microwave \cite{QC_WALTHERTRAP99} domains.

In the intermediate regime, for mesoscopic fields with an average photon number 
$\overline{n}\sim$ a few tens, the
amplitude of the Rabi oscillation is predicted to collapse and revive 
periodically \cite{QC_EBERLYREVIVE80}. The collapse is usually attributed to the
fluctuation of the field amplitude, and the revival to the graininess 
of the photon number. Entanglement between the atom and the field plays also
an important role in the process \cite{QC_BANACLOCHE90,QC_BANACLOCHE91,QC_KNIGHTREVIVE92,QC_KNIGHT95}. 
Due to the spread of the Rabi frequencies
corresponding to different photon numbers, the atom gets entangled 
with the field in a quantum superposition of two coherent
components rotating in opposite directions in phase space. 
These components are correlated to two different atomic state
superpositions. The Rabi oscillation appears as a quantum 
interference effect between the amplitudes associated to these atomic
superpositions. Its collapse is a direct consequence of complementarity. When 
the field is split into two orthogonal components with different phases,
it stores an unambiguous information about these atomic states, thus 
destroying the interference between them. The Rabi oscillation
revives when this information is erased, as the two components of the 
field periodically merge together. 

Evidence of the collapse
and revival of the Rabi oscillation has been obtained in 
CQED experiments \cite{QC_WALTHERREVIVAL87,ENS_QRABI}
with fields containing about one photon. Collapses and revivals
have also been observed in ion-traps with $\sim 3$ ion vibration quanta replacing
the photons \cite{ION_WINELANDRABI96}. These experiments focused on the
atomic evolution in a regime where only a few photons/phonons were involved.

We describe here an experiment detecting 
the evolution of a mesoscopic field containing an average photon number of up to about 40 and 
interacting with a single atom in a high $Q$ microwave cavity. The phase splitting 
of the field into two components rotating
in opposite directions in phase space is observed by a homodyne 
method \cite{ENS_GATE,ENS_WIGNER}. A single atom splits
the field  into phase components separated by up to 90 degrees, 
entangled with atomic states having initially opposite 
expectation values of their dipole. This
experiment demonstrates that an atom leaves its imprint on a 
mesoscopic field, opening the way to applications in CQED.

A two-level atom ($e,g$), initially in $g$,
and a single resonant field mode containing $n+1$ photons at $t=0$,
evolve into the state at time $t$ 
$|{\psi}_n (t) \rangle =\cos(\Omega
\sqrt{n+1}t/2)\\|g,n+1\rangle-i\sin(\Omega 
\sqrt{n+1}t/2)|e,n\rangle$.
This describes a reversible oscillation between an atom in $e$ with $n$ photons and
in $g$ with $n+1$ photons.
We are interested here, rather, in the
situation where the field is initially 
in a coherent state, superposition of number states $|n\rangle$, defined
as $|\alpha \rangle=e^{-|\alpha|^{2}/2}\sum_{n}(\alpha^{n}/\sqrt{n!})|n\rangle$
(where $\alpha$ is taken real without loss of generality).
The mean photon number is $\overline{n}=|\alpha|^2$. The 
photon number and phase fluctuations are $\Delta n =
\sqrt{\overline{n}}=|\alpha|$ and $\Delta \phi=1/\Delta n= 
1/\sqrt{\overline{n}}$ respectively. The atom-field state at time
$t$ can be derived exactly by superposition of the partial states 
$|{\psi}_n (t)\rangle$ associated to
the $|n\rangle$ states in the expansion of $|\alpha\rangle$. This solution 
describes a beating between
probabilities evolving at the incommensurate frequencies $\Omega
\sqrt{n+1}$.  

A more transparent
expression for the system's state is obtained for large $\overline{n}$, 
where the relative fluctuation of the photon numbers
is small \cite{QC_BANACLOCHE90,QC_BANACLOCHE91,QC_KNIGHTREVIVE92,QC_KNIGHT95}. It is then legitimate to replace in the exact expression 
$\sqrt{n+1}$ by $\sqrt{n}+1/(2\sqrt{\overline{n}})$ and
to develop then all functions of $n$ around $\overline{n}$ up to second order 
in $n-\overline{n}$. The system's state in the interaction 
picture then becomes at time $t$:
\begin{eqnarray}
|{\psi} (t) \rangle &\simeq& \frac{1}{\sqrt{2}}\Big[e^{-
i\Omega\sqrt{\overline{n}}t/2}|{\alpha}^{+}(t)\rangle
|{\phi}_{a}^{+}(t)\rangle\nonumber\\
&&-e^{i\Omega\sqrt{\overline{n}}t/2}|{\alpha}^{-}(t)\rangle
|{\phi}_{a}^{-}(t)\rangle\Big]\ ,
\label{EQ_EVOLUTION}
\end{eqnarray}
where the field states $|{\alpha}^{\pm}(t)\rangle$ and 
atomic states $|{\phi}_{a}^{\pm}(t)\rangle$ are:
\begin{eqnarray}
|{\alpha}^{\pm}(t)\rangle&=& 
e^{-|\alpha|^{2}/2}e^{\pm i\Omega 
\sqrt{\overline{n}}t/4}\sum_{n}e^{\pm
i\Omega(n-\overline{n})^2t/(16\overline{n}^{3/2})}\times\nonumber\\
&&\qquad \times\frac{(\alpha e^{\mp 
i\Omega t/4\sqrt{\overline{n}}})^n}{\sqrt{n!}}|n\rangle\label{EQ_FIELD}\\
|{\phi}_{a}^{\pm}(t)\rangle&=& \frac{1}{\sqrt{2}}[e^{\mp
i\Omega t/4\sqrt{\overline{n}}}|e\rangle\pm|g\rangle]\label{EQ_ATOM}\ .
\end{eqnarray}
From Eq.(1), we derive the probability $P_{g}(t)$ to 
detect at time $t$ the atom in level $g$:
\begin{equation}
P_{g}(t)=\frac{1}{2}
\big[1+\Re(e^{-i\Omega\sqrt{\overline{n}}t}\langle{\alpha}^{-}(t)| 
{\alpha}^{+}(t)\rangle)\big]
\label{EQ_RABI}
\end{equation}

Eqs.(\ref{EQ_EVOLUTION}-\ref{EQ_RABI}) yield a synthetic view of the evolution of
a mesoscopic field coupled to a single atom. At time
$t=0$, Eqs. (\ref{EQ_FIELD}) and (\ref{EQ_ATOM}) reduce to 
$|{\alpha}^{\pm}(0)\rangle=|\alpha\rangle$ and
$|{\phi}_{a}^{\pm}(0)\rangle=(|e\rangle\pm| 
g\rangle)/\sqrt{2}$. The initial atomic state $|g\rangle$
appears as the superposition of two orthogonal ``dipole states" with a 
non-zero mean dipole either
in phase [$(|e\rangle +|g\rangle)/\sqrt{2}$] or $\pi$-out of phase [$(|e\rangle 
-|g\rangle)/\sqrt{2}$] with the field. As time proceeds, the phases of
these two dipole states rotate in opposite directions in phase space 
at frequency $\pm \Omega/(4\sqrt{\overline{n}})$ [Eq.(\ref{EQ_ATOM})]. 
Simultaneously, the field splits into two quasi-coherent components
with phases $\Phi^\pm=\pm\Omega t/4\sqrt{\overline{n}}$, each 
remaining locked in phase or $\pi$-out of 
phase with the two rotating atomic ``dipole states" [Eqs.(\ref{EQ_EVOLUTION}) and 
(\ref{EQ_FIELD})]. The field components have their phases not only
shifted, but also spread out. The amplitude factor $[\alpha \exp(\mp i 
\Omega t/(4\sqrt{\overline{n}})]^n$ in Eq.(\ref{EQ_FIELD}) 
accounts for the phase drift, while the factor 
 whose exponent 
is quadratic in $n$ is responsible for phase spreading.

As a result of the
atom-field evolution [Eq.(\ref{EQ_EVOLUTION})], the two systems are 
generally entangled. The probability amplitudes associated with
the two components of this entangled state evolve at frequency
$\pm\Omega\sqrt{\overline{n}}/2$, $2\overline{n}$ times larger than the 
drift frequency of the atomic and field
phases. The beating between these two fast oscillating amplitudes 
leads to the Rabi oscillation [Eq.(\ref{EQ_RABI})],
which appears as an atomic interference effect.  The 
envelope of the Rabi oscillation is the overlap between
the two field components, a clear manifestation of complementarity.
Note that this discussion is also valid, within minor changes, when 
the atom is initially in $e$.

The mesoscopic nature of the field is essential. If the photon number 
is microscopic
($\overline{n}\leq 10)$, the approximations leading to 
Eqs.(\ref{EQ_EVOLUTION}-\ref{EQ_RABI}) are not 
valid. The classical 
limit corresponds conversely to
$\Omega\rightarrow 0$, $\overline{n}\rightarrow\infty$, with 
$\Omega\sqrt{\overline{n}}$ constant. The graininess of the photon 
number is then washed out, with collapse and revival times 
pushed out to infinity.

Our CQED set-up, sketched in Fig.1(a), is described in detail in \cite{ENS_RMP}.
$^{85}$Rb atoms, effusing from oven $O$, are velocity-selected by 
laser optical pumping and prepared in zone $B$ in the
circular Rydberg state with principal quantum number $51$ (level $e$) 
or $50$ (level $g$) by a combination of laser and
radiofrequency excitations. The atomic preparation is pulsed, so that 
the position of each atom is known along the beam within
$\pm 1$ mm. The atoms cross, one at a time, the cavity $C$ sustaining a 
Gaussian field mode (waist $w=6$ mm) exactly resonant with
the $e\rightarrow g$ transition at $51.1$ GHz. The cavity, cooled to 
0.6 K, is made up of two superconducting niobium mirrors. The
vacuum Rabi frequency 
is $\Omega =3.10^5$ s$^{-1}$. The atom and field relaxation
times are $T_{a}=30$ ms and $T_{cav}=850\ \mu$s, corresponding to $\Omega 
T_{cav}=250$, fulfilling the strong coupling regime condition. 
The atomic velocity is chosen from two values, $v_{a}=335$ m/s and
$v_{b}=200$ m/s, corresponding to an effective atom-cavity 
interaction time $t_i=t_{a}=\sqrt{\pi}w/v_a= 32\ \mu$s or 
$t_i=t_{b}=\sqrt{\pi}w/v_b=53\ \mu$s. 
The atoms are detected after $C$ by a field-ionization detector 
$D$ (quantum efficiency $\sim 70\%$)
discriminating $e$ and $g$.

A coherent field, produced by a pulsed (duration $23\ \mu$s) microwave source 
$S$, is injected in $C$ through a small hole
in a mirror. The 
average photon number $\overline{n}$ is
controlled with attenuators. An independent
measurement of $\overline{n}$ (with a precision of $\pm 10\%$) is realized 
by detecting the light shift
produced by the field on a Rydberg atom \cite{ENS_FROMLAMB}.

The timing of the experimental sequence is shown on the space-time 
diagram of Fig. 1(b). The preparation box $B$,
cavity $C$ and detector $D$ are represented, from left to right, by 
vertical grey bands. The cavity
is initially in the vacuum state. A first atom $A_{1}$, prepared in level $e$ 
or $g$ with velocity $v_{a}$ or $v_{b}$ is sent across the
set-up (lower diagonal line). Just before $A_{1}$ reaches $C$, a 
coherent field $F_{1}$ is injected into the
mode (lower horizontal line). The atom then crosses $C$. As soon as 
it exits the mode, a probe field $F_{2}$ (upper horizontal line) is injected, with the same amplitude as 
$F_1$, and a relative phase $\phi+\pi$. The $F_{1}$ and $F_{2}$ fields add in $C$ 
and their sum is read out by a second, probe
atom $A_{2}$ prepared in $g$, reaching $C$ just after the injection of $F_{2}$
(upper diagonal line).

This procedure amounts to a homodyne phase sensitive detection \cite{ENS_GATE,ENS_WIGNER}. 
Atom $A_{2}$ absorbs the final 
field in $C$, ending up in
$e$ with a large probability when $\phi$ is such that there is one photon
or more in $C$. By repeating the
sequence many times, we construct a signal $S_g(\phi)$ equal to the 
probability versus $\phi$ that $A_2$ remains in
$g$. This signal exhibits peaks
revealing the final phase pattern of the field in $C$ after interaction
with $A_1$. The
$A_{2}$ signal is observed in coincidence with the detection of $A_{1}$.

Fig. 2 shows $S_g(\phi)$ for the two $A_{1}-F_{1}$ interaction times 
$t_{a}$ [2(a)] and $t_{b}$ [2(b)]. The signals are plotted versus 
$\phi$ for different mean photon numbers
$\overline{n}$ in the range $\overline{n} = 15$ to $36$. Fig. 2(b) also shows the 
signal without $A_{1}$, for
$\overline{n}= 29$ photons, whose peak is centered around phase origin $\phi=0$. The
splitting by a single atom of the field into two symmetrical 
components with different phases is clearly visible in the other
recordings. For a given interaction time, the splitting decreases 
with field amplitude and, for a given $\overline{n}$, the
separation increases with time. Fig. 2(c) summarizes the results by 
plotting the phases of the two components as a function of
the dimensionless parameter $\Phi^+=\Omega t/4\sqrt{\overline{n}}$.
The dotted line corresponds to the phases $\Phi^\pm$ predicted
by Eq.(\ref{EQ_EVOLUTION}). The solid line results from a numerical simulation
solving the exact field equation of motion and
taking into account cavity damping.  The agreement 
between the experimental points and the solid
lines is very good. 
The maximum phase splitting observed, for
$\overline{n} = 15$ photons and $t_{i}=  52\ \mu$s, is 90 degrees.

The inset in Fig. 2(c) shows, for $\overline{n}=36$ and $t_i=t_a$ 
the Wigner function $W(\beta_x+i\beta_y)$ of the field in the cavity. 
It results from the
explicit numerical simulation. The field state is computed for an atom
found in $g$,~48 $\mu$s after its crossing of the cavity axis. This
$W$ function clearly 
exhibits the two separate field components and interferences which are a signature 
of a ``Schr\"odinger cat" coherence. The 
square of the distance in phase space between the two components,
$d^2=4\overline{n}sin^2(\Omega t/4\sqrt{\overline{n}})$, is a measure of the 
mesoscopic character of this superposition. In the
range of $\overline{n}$ we have explored, $d^2$ is nearly constant versus 
$\overline{n}$, equal to
$\sim 20$  for $t_i=t_{a}$ and $\sim 40$ for $t_i=t_{b}$. 
These figures are to be compared with the
size of the Schr\"odinger cats realized in our earlier CQED experiment ($d^2
\leq 8$) \cite{ENS_CAT}. Larger ``cat" states are obtained here, taking advantage of a higher-$Q$ 
cavity and of the
resonant atom-field interaction, which achieves a larger phase 
splitting than the dispersive coupling used in \cite{ENS_CAT}. The
theoretical decoherence time of the final cat state, $2T_{cav}/d^2$, is 
$\sim 43\ \mu$s for $t_i=t_{b}$, 
meaning that the superposition loses its coherence as 
fast as it is generated. The situation is better for
the smaller ``cats" prepared faster ($t_i=t_{a}$), which have a 
decoherence time $\sim 85\ \mu$s.

We have also checked the 
correlation between the atomic state and the field
phase by selectively preparing $|\phi_a^+(0)\rangle$ or 
$|\phi_a^-(0)\rangle$ at the beginning of the interaction, 
within a time short enough so that the slow phase 
drift of the atom and field states can be neglected. To prepare $|\phi_a^+(0)\rangle$,
the atom, initially in $g$, first performs a $\pi/2$ Rabi pulse according to the transformation
$g\rightarrow [e^{-i\pi/4}|\phi_a^+(0)\rangle-e^{i\pi/4}|\phi_a^-(0)\rangle]/\sqrt2$.  
The atom is then detuned by Stark effect with respect to the cavity, 
during a time much shorter than the
Rabi period. A pulse of electric field is applied between the cavity mirrors, whose 
effect is to shift by $\pi/2$ the relative phase of the $e$ and $g$ states \cite{ENS_RMP}. 
The sequence of Rabi and Stark pulses transforms the initial $g$ state,
superposition of the interfering $|\phi_a^+(0)\rangle$ and 
$|\phi_a^-(0)\rangle$ states,
into $|\phi_a^+(0)\rangle$ alone. The system ends up in the slowly evolving 
quasi-stationary state described by the first term in the
right hand side of Eq.(\ref{EQ_EVOLUTION}). The atom and the field 
subsequently drift in phase in only one 
direction. We observe that the Rabi oscillation is frozen from then on.
A homodyne
measurement of the field phase after the atom exits from $C$ then reveals, 
as expected, only a single phase-shifted field component (open circles in Fig. 3). 
Similarly, we have prepared $|\phi_a^-(0)\rangle$
by applying the same Rabi and Stark switching pulse sequence starting from 
level $e$. This state
couples to the other component of the field as revealed by the subsequent homodyne
detection (solid squares in Fig. 3). This experiment clearly demonstrates
correlations between the atomic state and the field phase.

Further tests of the quantum coherence in this system are under way. 
After collapse of the Rabi oscillation, 
we apply to the atom, at a time $T$, a Stark pulse 
switching the signs of the quantum
amplitudes associated to $e$ and $g$. According to Eq.(\ref{EQ_EVOLUTION}), this pulse 
suddenly exchanges the atomic states
correlated to the $|{\alpha}^{+}(t)\rangle$ and $| 
{\alpha}^{-}(t)\rangle$ field components. The atom-field
coupling resumes afterwards, reversing the sign of rotation of the two field
components. At time $2T$, the two field states are back in phase and 
the Rabi oscillation revives, revealing the coherent nature of
the atom/cavity state. This
induced revival, similar to a spin-echo, will be described in a forthcoming
paper. 

We have shown that a single atom leaves a quantum imprint on a 
mesoscopic field made of several tens of photons.
This experiment illustrates a striking feature of 
quantum-classical correspondence.
The $1/\sqrt{\overline n}$ dependence of $\Phi^+$ shows that
the bigger the mesoscopic system is, the longer is the time required for the atom to
pass its ``quantumness" to the field.
At the classical limit this time 
goes to infinity. Using better cavities and slower atoms
should allow us to prepare mesoscopic state superpositions with $\overline{n}$ 
in the range of a few hundred. We plan to study
the properties of these states and their decoherence. At a practical 
level, the large phase shifts produced by a
single atom on a mesoscopic field could be used to develop new atomic 
detection schemes. The information carried by a single atom
can be transferred to a field containing a large number of photons, 
which can be subsequently read out
by a sample made of many probe atoms.  This amplifying effect opens 
promising perspectives for a $\sim100\%$ atomic detection
efficiency in microwave CQED experiments.

\begin{acknowledgments}
 Laboratoire Kastler Brossel is a laboratory of
Universit\'e Pierre 
et Marie Curie and ENS, associated to CNRS (UMR 8552). We acknowledge support 
of the European Community, of the Japan Science and Technology corporation 
(International Cooperative Research Project~: Quantum Entanglement). 
\end{acknowledgments}

\begin{figure}
\includegraphics[width=5cm,height=5cm]{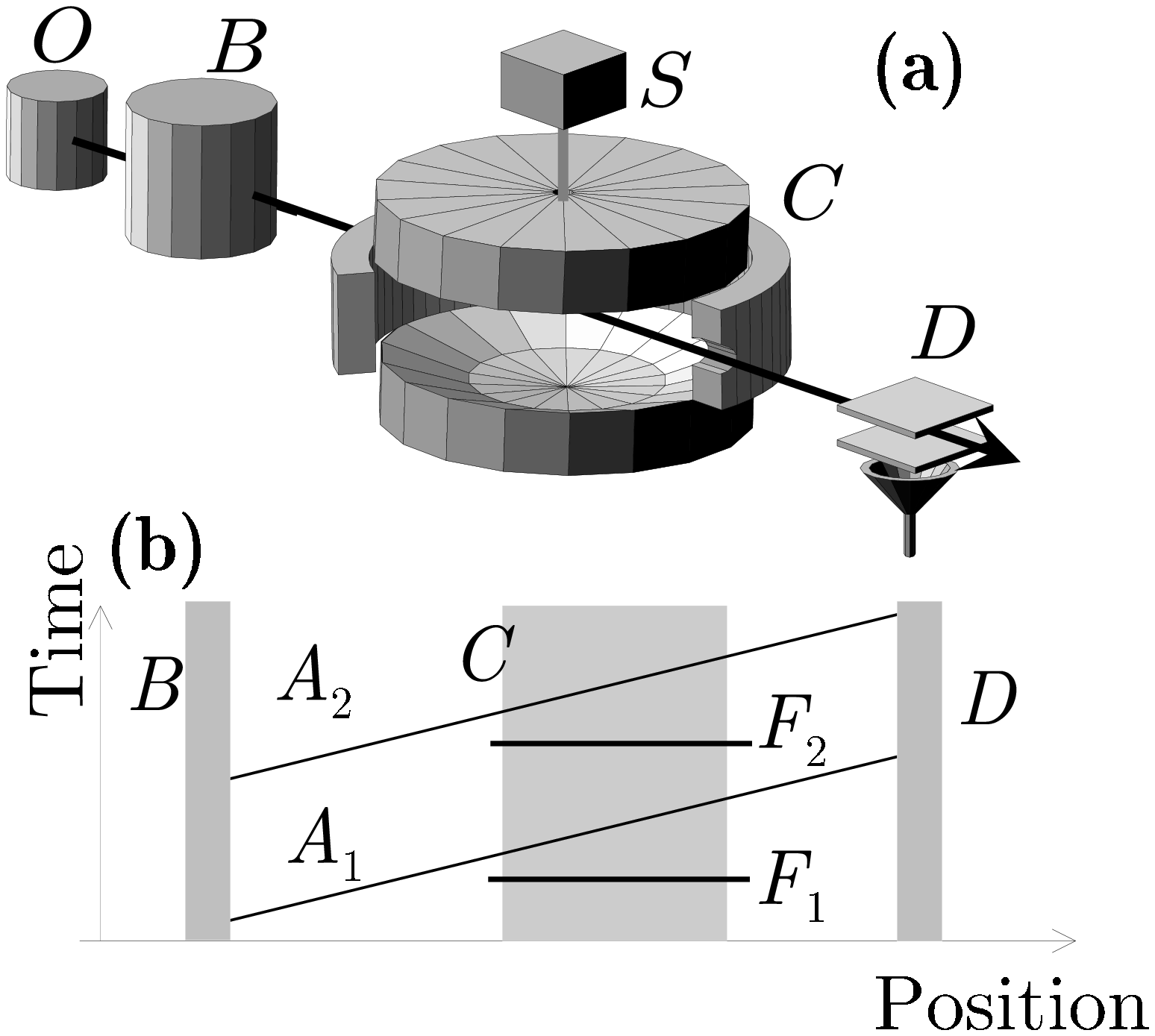}
\caption{(a) Experimental apparatus. (b) Timing of the 
experiment.}
\label{FIG1}
\end{figure}

\begin{figure}
\includegraphics[width=8cm,height=10cm]{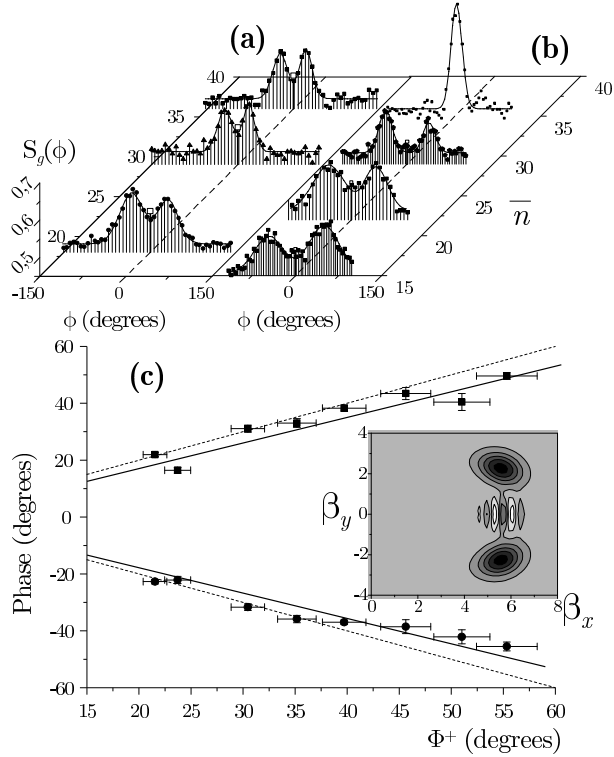}
\caption{(a) Field phase distribution $S_g(\phi)$ for three $\overline n$ values (18, 29 and 36) and $t_i=t_a=32\ \mu$s. The points are experimental and the curves fits on a sum of Gaussians. The thick vertical line indicates phase reference. (b) Field phase distribution for $\overline{n}=$15, 22 and 29 and $t_i=t_b=53\ \mu$s. Upper curve: phase reference (no atom $A_1$ and $\overline n=29$; for clarity, this curve has been translated along the $\overline n$ axis). (c) Phases of the two field components versus $\Phi^+=\Omega t/4\sqrt{\overline{n}}$. Dotted and solid lines are theoretical (see text). Inset: computed cavity field Wigner function $W(\beta_x+i\beta_y)$ for $\overline{n}=36$ and $t_i=t_a$.}
\label{FIG2}
\end{figure}

\begin{figure}
\includegraphics[width=8cm,height=4cm]{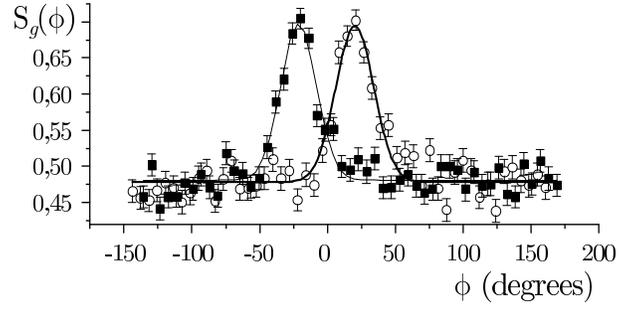}
\caption{Field phase distributions $S_g(\phi)$ following
preparation of atomic states $|\phi_a^\pm(0)\rangle$ by combination
of Rabi and Stark pulses ($\overline n=27$ and $t_i=t_a$). Open circles: 
preparation of $|\phi_a^+(0)\rangle$. Solid squares: preparation of 
$|\phi_a^-(0)\rangle$. Solid lines are gaussian fits.}
\label{FIG3}
\end{figure}

\end{document}